\documentclass[
    ,final            % use final for the camera ready runs
  ,numberedheadings % uncomment this option for numbered sections
  ]
  {aipproc}

\layoutstyle{8x11single}

\begin{document}

\title{The Correlation of Spectral Lag Evolution with Prompt Optical Emission in GRB 080319B\footnote{Contributed to the Proceedings of the Sixth Huntsville GRB Symposium. Edited by C.A. Meegan, N. Gehrels, and C. Kouveliotou.}}

%Replace this text with PACS numbers; choose from this list: http://www.aip..org/pacs/index.html
%95.75.Wx Time Series Analysis, Time Variability
%95.85.Kr Astronomical Observations (Visible)
%95.85.Pw Astronomical Observations (gamma-ray)
%98.70.Rz Unidentified sources of radiation outside of the solar system (gamma-ray sources; gamma-ray bursts)

\author{Michael Stamatikos\footnote{Correspondence to Michael.Stamatikos-1@nasa.gov.}$\;$}{
  address={Astroparticle Physics Laboratory, Code 661, NASA/Goddard Space Flight Center, Greenbelt, MD 20771 USA}
}

\author{Tilan N. Ukwatta}{
  address={The George Washington University, Washington, DC 20052 USA}
}

\author{Takanori Sakamoto}{
  address={CRESST, University of Maryland, Baltimore County, Baltimore, MD 21250 USA}
  ,altaddress={Astroparticle Physics Laboratory, Code 661, NASA/Goddard Space Flight Center, Greenbelt, MD 20771 USA}
}

\author{Kalvir S. Dhuga}{
  address={The George Washington University, Washington, DC 20052 USA}
}

\author{Kenji Toma}{
  address={Pennsylvania State University, University Park, PA 16802 USA}
}

\author{Asaf Pe'er}{
  address={Space Telescope Science Institute, Baltimore, MD 21218 USA}
}

\author{Peter M\'{e}sz\'{a}ros}{
  address={Pennsylvania State University, University Park, PA 16802 USA}
}

\author{David L. Band}{
  address={CRESST, University of Maryland, Baltimore County, Baltimore, MD 21250 USA}
  ,altaddress={Astroparticle Physics Laboratory, Code 661, NASA/Goddard Space Flight Center, Greenbelt, MD 20771 USA}
}

\author{Jay P. Norris}{
  address={Space Science Division, NASA/Ames Research Center, Moffett Field, CA 94035-1000 USA}
}

\author{Scott D. Barthelmy}{
  address={Astroparticle Physics Laboratory, Code 661, NASA/Goddard Space Flight Center, Greenbelt, MD 20771 USA}
}

\author{Neil Gehrels}{
  address={Astroparticle Physics Laboratory, Code 661, NASA/Goddard Space Flight Center, Greenbelt, MD 20771 USA}
}

\begin{abstract}
We report on observations of correlated behavior between the prompt $\gamma$-ray and optical emission from GRB 080319B, which confirm that (i) they occurred within the same astrophysical source region and (ii) their respective radiation mechanisms were dynamically coupled. Our results, based upon a new cross-correlation function (CCF) methodology for determining the \emph{time-resolved} spectral lag, are summarized as follows. First, the evolution in the arrival offset of prompt $\gamma$-ray photon counts between Swift-BAT 15-25 keV and 50-100 keV energy bands \emph{(intrinsic $\gamma$-ray spectral lag)} appears to be anti-correlated with the arrival offset between prompt 15-350 keV $\gamma$-rays and the optical emission observed by TORTORA \emph{(extrinsic optical/$\gamma$-ray spectral lag)}, thus effectively partitioning the burst into two main episodes at $\sim T+28\pm2$ sec.
Second, the rise and decline of prompt optical emission at $\sim T+10\pm1$ sec and $\sim T+50\pm1$ sec, respectively, both coincide with discontinuities in the hard to soft evolution of the photon index for a power law fit to 15-150 keV Swift-BAT data at $\sim T+8\pm2$ sec and $\sim T+48\pm1$ sec. These spectral energy changes also coincide with intervals whose time-resolved spectral lag values are consistent with zero, at $\sim T+12\pm2$ sec and $\sim T+50\pm2$ sec. These results, which are robust across heuristic permutations of Swift-BAT energy channels and varying temporal bin resolution, have also been corroborated via independent analysis of Konus-Wind data. This potential discovery may provide the first observational evidence for an implicit connection between spectral lags and GRB emission mechanisms in the context of canonical fireball phenomenology. Future work includes exploring a subset of bursts with prompt optical emission to probe the unique or ubiquitous nature of this result.
\end{abstract}

\maketitle

%%%%%%%%%%%%%%%%%%%%%%%%%%%%%%%%%%%%%%%%%%%%
%% MAINMATTER
%%%%%%%%%%%%%%%%%%%%%%%%%%%%%%%%%%%%%%%%%%%%

\section{Introduction}

\begin{figure}
\includegraphics[height=.54\textheight]{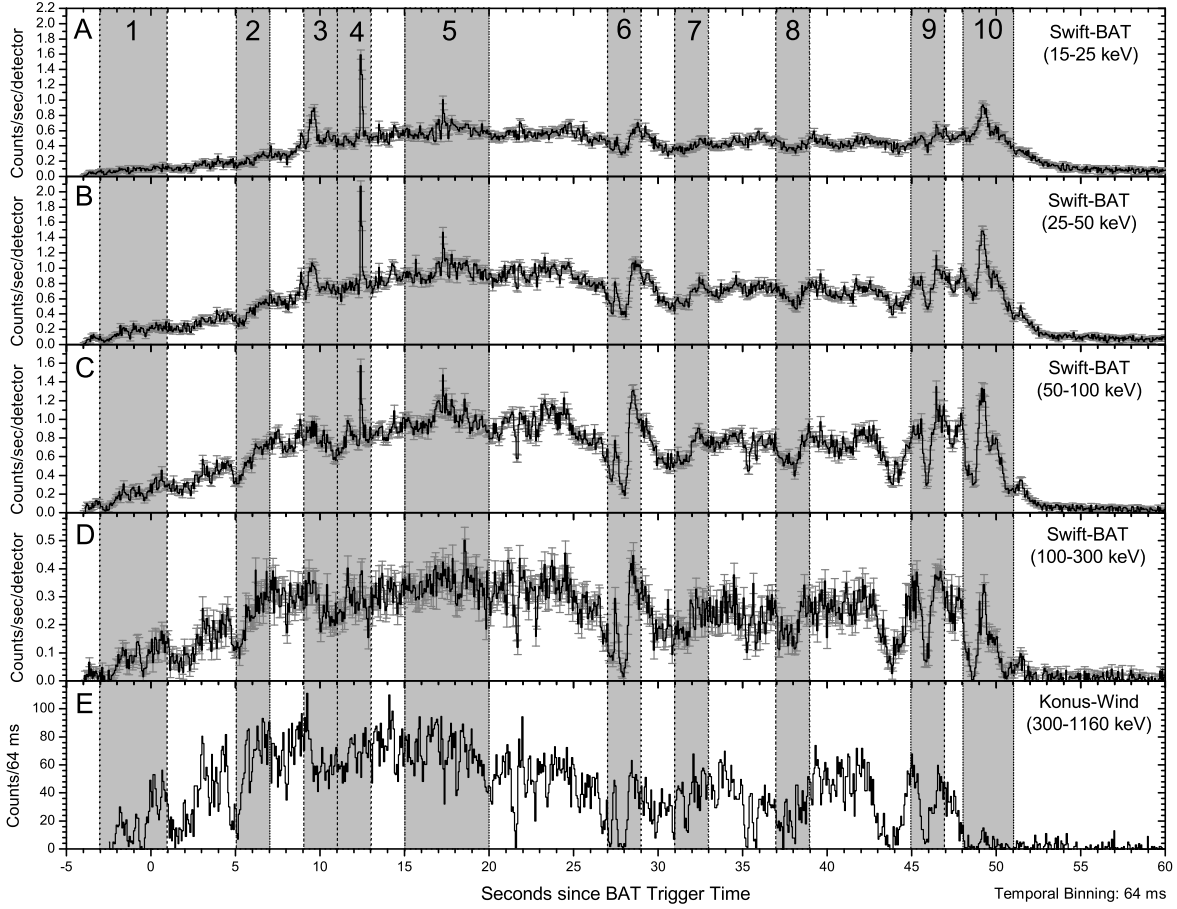}
\caption{64-ms binned light curves for Swift-BAT 15-25 keV (Panel A), 25-50 keV (Panel B), 50-100 keV (Panel C), 100-300 keV (Panel D) and Konus-Wind 300-1160 keV (Panel E) energy band passes. Gray regions indicate the sample periods for time resolved analysis. BAT error bars (Panels A-D) indicate 1$\sigma$ uncertainty. Konus-Wind data was taken from a supplement online \cite{Racusin:2008c}.}
\label{BAT_LC}
\end{figure}

Swift's unique dynamic response and spatial localization precision, in conjunction with correlative ground-based follow-up efforts, has resulted in the collaborative broad-band observations of GRB 080319B \cite{Racusin:2008c}. In the context of an analysis focused on confronting the lag-luminosity relation \cite{Norris:2000b} in the Swift era via the Burst Alert Telescope (BAT) \cite{Stamatikos:2008b}, a correlation was observed between the evolution of time-resolved spectral lag and the behavior of the extraordinarily well-sampled prompt optical emission light curve associated with GRB 080319B \cite{Stamatikos:2008}. In general, the spectral lag is determined via either a peak pulse fit \cite{Norris:2005b} or cross-correlation function (CCF) analysis \cite{Band:1997b}. Previous studies have reported on the variability of spectral lag throughout burst emission \cite{Chen:2005}, as well as its correlation to pulse evolution \cite{Hakkila:2008}. In this work, we develop a new method to calculate the time-resolved spectral lag of GRB 080319B, within 10 emission periods, as illustrated in Figure~\ref{BAT_LC}, via a modification to the traditional CCF approach. In this manner, we are able to explore for the first time the evolution and correlation of the intrinsic $\gamma$-ray spectral lag with prompt optical emission. We interpret these correlated behaviors as strong observational evidence that the prompt optical and $\gamma$-ray emission of GRB 080319B took place within the same astrophysical source region, with indications that their respective radiation mechanisms were dynamically coupled throughout the prompt phase of the burst.

\begin{figure}
\includegraphics[height=.44\textheight]{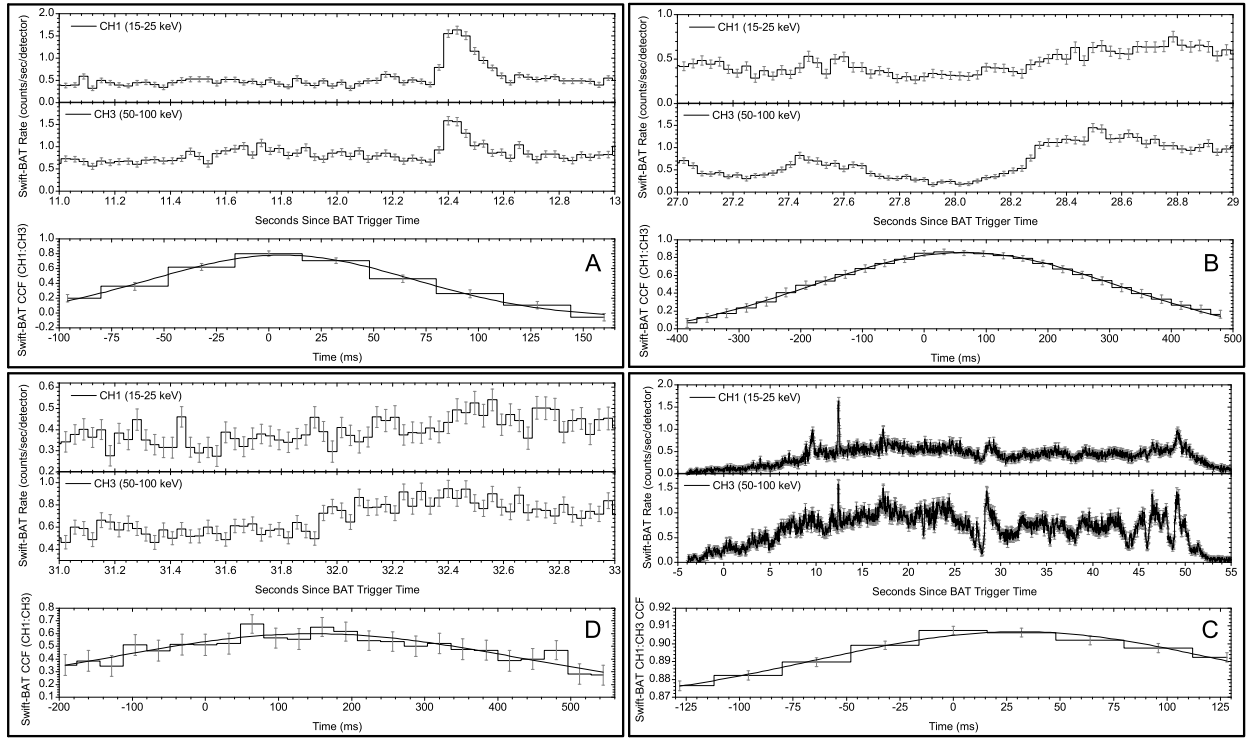}
\caption{Swift-BAT prompt $\gamma$-ray (32 ms bin) light curves for canonical energy channels 1 (15-25 keV) and 3 (50-100 keV), with corresponding CCF as a function of spectral lag time. The peak of a Gaussian fit (solid line) determined the time-resolved spectral lag. The results for sample time intervals 4, 6 and 7 (defined in Figure~\ref{BAT_LC}), include $6\pm4$ ms (\emph{\textbf{Panel A}}), $65\pm15$ ms (\emph{\textbf{Panel B}}) and $149\pm57$ ms (\emph{\textbf{Panel D}}), respectively. Note that the time-resolved spectral lag of a given interval (\emph{\textbf{Panels A, B \& D}}) can differ greatly from the time-averaged spectral lag of $29\pm4$ ms, as illustrated in \emph{\textbf{Panel C}}. All error bars denote $1\sigma$ uncertainty.}
\label{Period_6}
\end{figure}

\section{Methodology}

The Burst Advocate $\texttt{(BA)}$ script was used to generate BAT event data and quality maps, which were used to construct 32 ms binned\footnote{ A binning of 32 ms was used based upon a heuristic exploration of various bin resolutions (e.g. 4 ms, 8 ms, 16 ms, 32 ms, 64 ms, etc.). Once a spectral lag measurement was made, it was reproduced using light curves of several adjacent bin resolutions. In principle, the binning of the light curve should be smaller than the spectral lag in order to resolve the CCF peak.} (mask-weighted\footnote{Although one increases the signal-to-noise ratio (SNR) using non-mask-tagged (raw, non-background subtracted) light curves, artifacts either intrinsic to the detector or associated with the background may mimic signal pulses and thus be treated as such in a CCF analysis. Also, the background may not be trivial to model since it may be highly variable due to slewing. Hence, in order to be conservative and consistent with future studies, background subtracted, i.e. mask-tagged, light curves were used throughout the analysis.}, background-subtracted) light curves (temporal photon spectra) for $E_{A}=15-25$ keV and $E_{B}=50-100$ keV energy bands\footnote{Permutations of canonical BAT channel (1-4) light curve pairings were investigated. Ultimately BAT channels 1 (15-25 keV) and 3 (50-100 keV) were used in the analysis since they represented the largest energy differential with the highest count rate significance.}, via the $\texttt{BATBINEVT}$ analysis task. The resultant $\texttt{FITs}$ files were analyzed via a modified CCF method \cite{Band:1997b} in order to quantify the temporal correlation between the two series of GRB light curves in differing energy channels, i.e. $E_{A}$ and $E_{B}$, via the following:

\begin{equation}
CCF\left(t_{o},\nu_{E_{A}},\nu_{E_{B}}\right)=\frac{\langle\nu_{E_{A}}(t)\nu_{E_{B}}(t+t_{o})\rangle}{\sigma_{\nu_{E_{A}}}\sigma_{\nu_{E_{B}}}}\equiv CCF_{AB},
\label{CCF}
\end{equation}
where $\sigma_{\nu_{E_{n}}}=\langle\nu_{E_{n}}^{2}\rangle^{1/2}$. The CCF between $E_{A}$ and $E_{B}$ peaks at a given temporal offset $\left(t_{o}\right)$, known as the \emph{spectral lag} $\left(\tau_{AB}\right)$, which is defined as positive if the systematic shift in the arrival times of photon counts between pairs of light curves results in higher energy $\left(E_{B}\right)$ photons arriving \emph{before} those of lower energy $\left(E_{A}\right)$. The generic CCF is based upon the Pearson Correlation (Equation~\ref{CCF}). If the mean is subtracted\footnote{Mean subtraction is the formalism identical to that used by the \texttt{C\_CORRELATE} function of \texttt{IDL 6.2}. The formal definition may be found online (\texttt{http://c1.dmf.arm.gov/base/idl\_6.2/C\_CORRELATE.html}).}, then $v(t) = d(t) -\langle d(t)\rangle$. If the mean is not subtracted, then $v(t) = d(t) - b(t)$, where an attempt is made to remove background counts $b(t)$. Our tests indicated that both methods agreed when one calculates the \emph{time-averaged} spectral lag, i.e. the spectral lag determined over the entire duration of the GRB (Figure~\ref{Period_6}, Panel C). This included a test of subtracting the mean of a light curve that was nested within two large background intervals, which had the effect of reducing the mean and increasing the correlation amplitude. Note that if one takes an infinite interval, the mean would go to background, which fluctuates about zero, resulting in effectively not subtracting the mean in the first place, as prescribed for transient sources such as GRBs \cite{Band:1997b}.

However, when one tries to extract a \emph{time-resolved} spectral lag, i.e. the spectral lag over a segment of the light curve, then only the mean subtraction method consistently works. This is a consequence of edge effects introduced by the assumptions of a stationary versus a transient temporal signal characterization, which destroy/diminish the intrinsic lag. If one subtracts the mean, there is an assumption that the signal exists as a stationary wave throughout time, i.e. beyond the sampled time series. If one does not subtract the mean, then the assumption is that the signal is entirely contained within the time series. Both methods disregard data beyond the overlapped interval displaced via bin shifts.

Hence, for a bright GRB like 080319B, a non-mean subtracted time-resolved spectral lag analysis is dominated by the segment of the light curve described by a square wave, and the real light curve signal is treated as fluctuations. A mean-subtracted time-resolved spectral lag analysis will extract the spectral lag since it smoothes out the amplitude and focuses on the signal variance. Hence, we utilized the mean-subtracted definition of the CCF, since it is both more conservative and robust. Although it has been demonstrated that spectral lag variability is ubiquitous in GRBs \cite{Chen:2005,Hakkila:2008}, previous methods required relatively well-behaved sub pulses, such as those canonically described as FREDs (fast-rise, exponential decay). Our new methodology treats structure without regard to its functional form and affords greater flexibility for time-resolved spectral lag evolution studies, as illustrated in Figure~\ref{Period_6}, which contrasts sample periods 4 (Panel A), 6 (Panel B) and 7 (Panel D) with the time-averaged interval (Panel C) for GRB 080319B.

\begin{figure}
\includegraphics[height=.54\textheight]{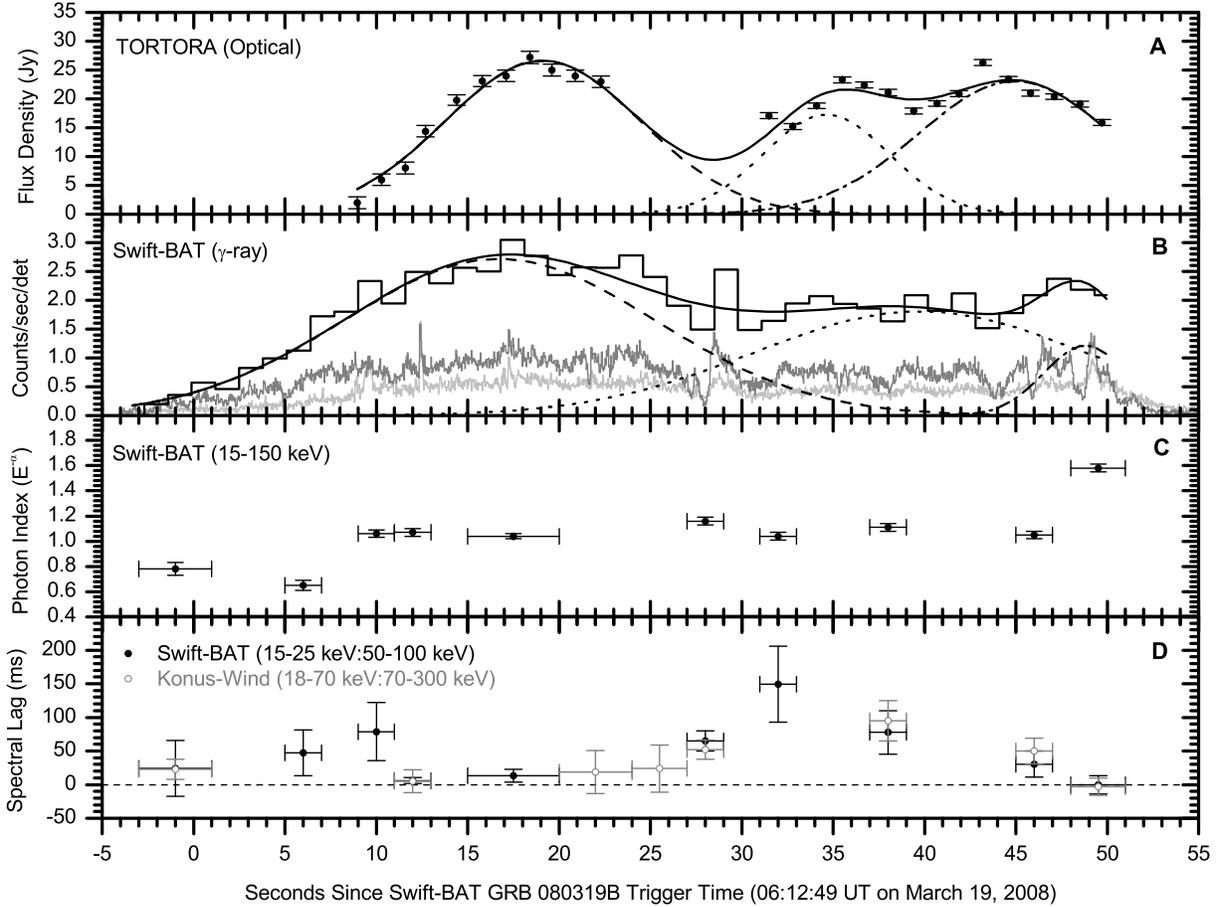}
\caption{\emph{\textbf{Panel A -}} TORTORA prompt optical flux density for $\sim$1.3 second exposure intervals (solid black circles) \cite{Racusin:2008c}. A cumulative fit (solid line) is based upon three Gaussian curves with peaks at $\sim$T+19 sec (dashed line), $\sim$T+34 sec (dotted line) and $\sim$T+45 sec (dash-dotted line). \emph{\textbf{Panel B -}} Swift-BAT prompt $\gamma$-ray light curves for (32 ms bin) canonical energy channels 1 (15-25 keV, light gray line), 3 (50-100 keV, dark gray line) and (1.3 second bin) 15-350 keV (black line). A cumulative fit (solid line) is based upon three Gaussian curves with peaks at $\sim$T+17 sec (dashed line), $\sim$T+40 sec (dotted line) and $\sim$T+49 sec (dash-dotted line). \emph{\textbf{Panel C -}} Photon index $(E^{-\alpha})$ for power law fit to 15-150 keV Swift-BAT data, with $1\sigma$ error bars. \emph{\textbf{Panel D -}} Time-resolved, intrinsic spectral lag between Swift-BAT 32 ms bin light curves for 15-25 keV and 50-100 keV energy bands (solid black circles) and Konus-Wind 64 ms bin light curves for 18-70 keV and 70-300 keV energy bands (open gray circles).}
\label{plot}
\end{figure}

In order to test the viability of an interval, a large known spectral lag was artificially introduced, which shifted one of the test light curve pairs. Intervals where the analysis was unable to recover the artificial spectral lag due to either low SNR or lack of structure were not used. The CCF method, regardless of definition, is sensitive to the SNR and, more importantly, to the shape of the region. Morphological tests using simple square, triangular and Gaussian shaped pulses have illustrated that both mean and non-mean subtracted CCF definitions failed to recover the artificial spectral lag that was obscured in regions where (i) only slopes were compared, (ii) SNR was low and/or (iii) error bars were large. This puts a fundamental limit on determining the time-resolved spectral lag for discrete intervals, which precluded an arbitrary systematic sampling of GRB 080319B, as illustrated in Figures~\ref{BAT_LC} and ~\ref{plot} (Panels C \& D). Hence, although our methodology is robust, it does require a selection on intervals with structured variability. This is not an issue for time-averaged results, since the burst is nested between periods of quiescence, as given in Figure~\ref{Period_6}, Panel C.

Considerable effort was given to ascribing a statistical significance to the spectral lag evolution via the determination of errors. Several methods were explored which included an approximation of confidence intervals, discrete CCF formulations and parameterizations accounting for the total height ($h$), half-width at half-maximum ($W_{c}$) and number of light curve bins ($n$) for a given CCF peak \cite{Gaskell:1987}, such that: $\sigma_{\tau}\approx\frac{0.75W_{c}}{1+h(n-2)^{1/2}}$.

Ultimately, we implemented a more intuitive approach, which featured a double Monte Carlo (MC) schema via canonical bootstrapping techniques \cite{Efron:1993}. In this manner, data-based simulations were used to estimate the standard error via the standard deviation of the bootstrapped light curve replications. For a given light curve bin $\left(LC_{bin}\pm\sigma_{LC_{bin}}\right)$, a simulated realization $\left(LC_{bin}^{\prime}\pm\sigma_{LC_{bin}}\right)$ of the count rate was constructed as follows: $LC_{bin}^{\prime}=LC_{bin}+\xi\sigma_{LC_{bin}}$. The random number seed $\left(\xi\right)$ was determined via the IDL \texttt{RANDOMN}\footnote{\texttt{RANDOMN} is described online at \texttt{http://idlastro.gsfc.nasa.gov/idl\_html\_help/RANDOMN.html}.} function, which is based upon the Box-Muller method and returns a normally-distributed, floating-point, pseudo-random number with a mean of zero and a standard deviation of one.

The error bars on the CCF function were determined via a primary MC simulation, which were constructed from 1000 realizations of light curve pairs, within the $1\sigma$ error bars of the data. The distribution of CCF values within a given bin defined its error bar, as illustrated in Figure~\ref{Period_6}. Once the error bars were fixed in this manner, a secondary MC simulation constructed a subsequent set of 1000 realizations of light curve pairs, within the $1\sigma$ error bars of the data. Each resulting CCF (with errors fixed from the primary MC simulation) was subjected to a Gaussian fit\footnote{Typically the CCF is asymmetric, with a distribution typically skewed to the right. This is due in part to the intrinsic hard to soft pulse width evolution. The CCF width is also affected by the smearing of multi-pulses within the segment. The skewness was not commonly minimized via the introduction of error bars from the primary MC simulation. In such cases, a region of local symmetry about the peak was selected using asymmetric endpoints. For extreme cases, a cubic function may be used to guide the functional fit in order to accommodate the skewness.}
\begin{equation}
y=Ae^{-\left(x-x_{c}\right)^{2}/\left(2w^{2}\right)},
\label{Gaussian}
\end{equation}
whose peak was extracted via the \texttt{IDL} function \texttt{MPFITPEAK\footnote{See \texttt{http://purl.com/net/mpfit}, for more information on \texttt{MPFIT}.}}. The peak fit to the data was used to determine the spectral lag value, while the numerical standard deviation of the distribution of simulated peaks was used to determine the error, which was in agreement with comparisons made with the cumulative fraction at 68\% and CCF shape $\left(\sigma_{\tau}\right)$, to within a factor of $\sim0.80\pm0.26$. We interpret this as the $1\sigma$ spectral lag error\footnote{We note that taking the peak fit error to the data resulted in values that were smaller by a factor of $\sim2.4\pm0.80$.}, as illustrated in Figure~\ref{plot}, Panel D.

The primary MC simulation optimized the CCF value per bin and provided a straight forward generation of CCF error bars, which are typically omitted, without appealing to cryptic statistical arguments or the assumptions of otherwise un-weighted fits. The secondary MC simulation optimized the CCF peak and simulated our data analysis. Since the CCF is normalized to unity, the correlation coefficient ascribes the confidence level of the correlation, which ranged from $\sim$40\%-90\%, with typical values of $\sim$60\%. Segments with correlation values $<\sim30\%$ were omitted.

In general, spectral lag calculations are sensitive to a series of selection parameters that include the energy band pass of each comparative light curve, temporal bin resolution and emission interval. The CCF and pulse fit methods are both based upon a functional expression for the peak fit (Gaussian, cubic, etc.), while the former requires a bin shift range. A self-consistent comparison, which includes uncertainties, must account for these selection effects. Efforts to standardize the spectral lag analysis are ongoing and would facilitate direct comparative studies across detector sensitivities and energy band passes \cite{Stamatikos:2008b}.

\section{Results \& Discussion}

Our preliminary results are illustrated in Figure~\ref{plot}, Panels A-D. A major result from our analysis is that the first episode in spectral softening of the photon index $\left(\frac{dN_{\gamma}}{dE_{\gamma}}\propto E_{\gamma}^{-\alpha}\right)$ at $\sim T+8\pm2$, based upon time-resolved power law fits to 15-150 keV Swift-BAT data intervals, coincides with both the initial rise of the optical flux and first apparent peak in spectral lag at $\sim T+10\pm1$ sec, as illustrated in Figure~\ref{plot}, Panels A, C and D. This is consistent with independent Konus-Wind time-resolved spectral analyses of hardness ratios \cite{Racusin:2008c}. Although the onset of the afterglow may obscure the analysis of prompt emission for $t>\sim T+50$ sec, the second episode of spectral softening at $\sim T+48\pm1$ sec coincides with the apparent decline of the optical flux and zero spectral lag at $\sim T+50\pm2$ sec, which is also independently corroborated via Konus-Wind time-resolved analyses \cite{Racusin:2008c}, as illustrated in Figure~\ref{plot}, Panels A, C and D.

The generic agreement of the overall temporal coincidence and morphology between the prompt $\gamma$-ray and optical light curves (Figure~\ref{plot}, Panels A \& B) in conjunction with the observed spectral and temporal evolution (Figure~\ref{plot}, Panels C \& D), suggests that they arose from a common source region (cf. \cite{Zou:2008}), in agreement with the double-jet hypothesis where both emissions occurred within the narrow jet \cite{Racusin:2008c}. However, separate radiation mechanisms were most likely responsible since the extrapolated $\gamma$-ray flux density to the optical band was deficient by $\sim$4 orders of magnitude when compared to observation \cite{Racusin:2008c,Kumar:2008,Yu:2008}. The steep rise/decline, short duration and lack of increasing pulse width of the prompt optical emission disfavor external forward/reverse shocks. Hence, internal shocks have been suggested as the source region with synchrotron emission responsible for the optical and inverse Compton scattering/synchrotron self Compton for the $\gamma$-rays, with associated GeV photon emission \cite{Racusin:2008c,Kumar:2008}. Alternatively, it has been suggested that non-relativistic forward internal shocks generated the prompt optical emission, while relativistic reverse internal shocks generated the prompt $\gamma$-rays, with sub-GeV/MeV photon emission \cite{Yu:2008}. Such high energy emission may be tested on future GRBs using Fermi (formally known as GLAST) via joint Swift-BAT analyses \cite{Stamatikos:2008c}.

Furthermore, there are several observations that lead to effectively separating the burst's duration into two main episodes partitioned roughly at the midpoint of $\sim T+28\pm2$ sec. The first is that the (bimodal) evolution of the intrinsic $\gamma$-ray spectral lag increases at t $>\sim T+28\pm2$ sec, which appears to be anti-correlated with the extrinsic (optical/$\gamma$-ray) spectral lag, which is of the order of a few seconds when t < $\sim T+28\pm2$ sec (Panels A \& B), as observed via the smoothed Gaussian fits, as illustrated in Figure~\ref{plot}, Panels A \& B. Beyond this common midpoint, the optical and $\gamma$-rays do not correlate as well or at least are ambiguously correlated, i.e. extrinsic (optical/$\gamma$-ray) spectral lag is either zero or negative at later times. The ambiguity is due to apparent peak misalignment. Hence, the intrinsic time-resolved $\gamma$-ray spectral lag is maximum at t $>\sim T+28\pm2$ sec, while the extrinsic time-resolved (optical/$\gamma$-ray) spectral lag is maximum at t $<\sim T+28\pm2$ sec. In addition, an independent analysis of BAT 15-150 keV light curves has revealed that the characteristic variability timescale of GRB 080319B was $\sim$100 ms for t < $\sim T$+28 sec and $\sim$1 sec for t > $\sim T$+28 sec \cite{Margutti:2008}. Furthermore, independent time-resolved spectral analysis of Konus-Wind data illustrated that peak energy decreased from $751\pm26$ keV to $537\pm28$ keV at $\sim T+24\pm2$ sec \cite{Racusin:2008c}.

In conclusion, we interpret these correlated behaviors as strong observational evidence that the prompt optical and $\gamma$-ray emission took place within the same astrophysical source region with dynamically coupled radiation mechanisms, which until now has only been conjecture \cite{Racusin:2008c,Kumar:2008,Yu:2008}. This potential discovery may provide the first observational evidence for an implicit connection between spectral lags and GRB emission mechanisms in the context of canonical fireball phenomenology. A full theoretical analysis of this result is currently in preparation. Future work includes an application of our methodology to observations of a subset of bursts with prompt optical emission to probe either the unique or ubiquitous nature of this result. Ultimately, understanding the mechanism(s) responsible for spectral lag may reveal a fundamental and unprecedented view from within the GRB fireball and its progenitor(s).

%%%%%%%%%%%%%%%%%%%%%%%%%%%%%%%%%%%%%%%%%%%%%%%%
%% BACKMATTER
%%%%%%%%%%%%%%%%%%%%%%%%%%%%%%%%%%%%%%%%%%%%%%%%

\begin{theacknowledgments}
The authors are grateful to Craig B. Markwardt for very fruitful discussions in regards to this analysis. M. Stamatikos is supported by an NPP Fellowship at NASA-GSFC administered by ORAU.
\end{theacknowledgments}



\bibliographystyle{aipproc}   % if natbib is available

\bibliography{Stamatikos_GRB_Huntsville_astroph}

\IfFileExists{\jobname.bbl}{}
 {\typeout{}
  \typeout{******************************************}
  \typeout{** Please run "bibtex \jobname" to optain}
  \typeout{** the bibliography and then re-run LaTeX}
  \typeout{** twice to fix the references!}
  \typeout{******************************************}
  \typeout{}
 }

\end{document}